# Nano-patterned back-reflector with engineered near-field/far-field light scattering for enhanced light trapping in silicon-based multi-junction solar cells

Andrea Cordaro[1,2,*], Ralph Müller[3], Stefan Tabernig[1,2], Nico Tucher[3], Patrick Schygulla[3], Oliver Höhn[3], Benedikt Bläsi[3], Albert Polman[1,2]

[1] Institute of Physics, University of Amsterdam, Science Park 904, 1098 XH Amsterdam, The Netherlands
[2] Center for Nanophotonics, NWO-Institute AMOLF, Science Park 104, 1098 XG Amsterdam, The Netherlands
[3] Fraunhofer ISE, Heidenhofstr. 2, 79110 Freiburg, Germany
*e-mail: a.cordaro@amolf.nl

**Multi-junction solar cells provide a path to overcome the efficiency limits of standard silicon solar cells by harvesting more efficiently a broader range of the solar spectrum. However, Si-based multi-junction architectures are hindered by incomplete harvesting in the near-infrared (near-IR) spectral range, as Si sub-cells have weak absorption close to the band gap. Here, we introduce an integrated near-field/far-field light trapping scheme to enhance the efficiency of silicon-based multi-junction solar cells in the near-IR range. To achieve this, we design a nanopatterned diffractive silver back-reflector featuring a scattering matrix that optimizes trapping of multiply-scattered light into a range of diffraction angles. We minimize reflection to the 0$^{th}$-order and parasitic plasmonic absorption in the silver by engineering destructive interference in the patterned back contact. Numerical and experimental assessment of the optimal design on the performance of single-junction Si TOPCon solar cells highlights an improved external quantum efficiency (EQE) over a planar back-reflector (+1.52 mA/cm$^2$). Nanopatterned metagrating back-reflectors are fabricated on GaInP/GaInAsP//Si two-terminal triple-junction solar cells via Substrate Conformal Imprint Lithography (SCIL) and characterized optically and electronically, demonstrating a power conversion efficiency improvement of +0.9%$_{abs}$ over the planar reference. Overall, our work demonstrates the potential of nanophotonic light trapping for enhancing the efficiency of silicon-based multi-junction solar cells, paving the way for more efficient and sustainable solar energy technologies.**

New solar power conversion architectures are increasingly important to achieve large-scale, efficient, and sustainable power generation to satisfy the growing energy needs of our society[1]. Many types of solar cells have reached high efficiencies. Yet, in all cell designs incomplete light absorption induces a loss in short-circuit current $J_{SC}$, while poor electronic carrier management (e.g. carrier recombination, parasitic resistance, etc.) results in reduced open-circuit voltage $V_{OC}$ or fill factor FF[2,3]. Although the majority of the current record cells still rely on single or double-pass absorption and exploit relatively thick absorber layers (>100μm for Si and >1 μm for thin-film architectures like GaAs, CdTe, etc.), smarter light managing strategies can lead to higher efficiencies but also a dramatic reduction of the absorber thickness. This, in turn, further implies numerous benefits including overall cost reduction, higher production throughput, flexible form factors for wearable technologies and vehicles, reduced weight, and more efficient charge extraction.

In the vast landscape of photovoltaic architectures, multi-junction solar cells are rapidly emerging[4–6]. With single-junction Si solar cells reaching power conversion efficiencies[7] close to the theoretical limit of 29.4%[8–10], major efficiency improvements can be obtained by combining multiple semiconductor materials with different band gaps in a multi-junction configuration so that light is absorbed more efficiently over a broad range in the solar spectrum, and a smaller fraction of the photon energy is lost due to thermalization[11]. Taking advantage of the maturity of Si solar cell technology, researchers have been exploring the best high band gap partners that can be coupled to state-of-the-art Si solar cells. However, while nanophotonic light management strategies[12–23] in thin-film PV can be designed by simulating the entire cell and optimizing absorption, Si-based multi-junction architectures require thicker cells and thus pose a unique design challenge as both ray optics and near-field nanoscale optics play a crucial combined role determining the overall performance of the cell. Due to the weak absorption in the Si sub-cell near the band gap (1000-1200 nm) part of the incoming solar light can escape from the front surface after being reflected at the bottom of the Si cell. When standard texturing of the Si wafer is not compatible with the device design, nanopatterned optical gratings can be used to steer light at angles at which total internal reflection (TIR) is occurring at the top interface[24]. Recent theoretical work[25,26] investigates the light-path enhancement induced by the grating period for light trapping in optically thick solar cells. However, a careful design and optimization of the grating at the nanoscale and at the same time an assessment of the resulting macroscale far-field light propagation through the entire cell is still missing.

Here, we introduce a near-field/far-field light scattering concept to design and implement a nano-structured back-reflector that optimizes the distribution of light scattering beyond the total internal reflection critical angle in a GaInP/GaInAsP//Si multi-junction cell with an efficiency above 35%. Specifically, we optimize the geometry of a hexagonal array of Ag nano-disks that are integrated with the Ag back contact, and show how pitch, radius, and height of the individual scatterers control the distribution of power over different diffraction orders, while at the same time minimizing plasmonic dissipation losses in the wavelength range of interest. We explain the physical mechanism governing light steering by the back reflector with an intuitive interference model and use a multiple-scattering matrix formalism to evaluate the near-bandgap light absorption in the Si bottom cell. We demonstrate experimentally, first on Si single-junction cells and then on full two-terminal triple-junction cells, an improved power conversion efficiency (+0.9%$_{abs}$), showing the benefits of this new light trapping design concept. Our design strategy applies to any Si-based multi-junction cell opening exciting opportunities for geometries that cannot support standard random texturing, including e.g. perovskite//Si tandem multi-junction solar cells.

**Theory and Design**

To demonstrate the concept of enhanced near-IR light harvesting in silicon-based multi-junction solar cells, we design and implement a nanostructured back-reflector at the bottom of a GaInP/GaInAsP//Si two-terminal triple-junction solar cell (Fig. 1a). These high-efficiency cells combine a thin (≈2 μm) III-V top tandem cell (GaInP/GaInAsP) to a thick bottom Si cell (≈300 μm) via direct wafer bonding and have reached an overall efficiency of 35.9%[24,27]. In this type of silicon-based multi-junction cells, weak absorption in the Si cell in the 1000-1200 nm wavelength range limits complete light harvesting. Furthermore, standard micro-texturing of the Si cell's frontside is not compatible with wafer bonding, while passivation of the p-type polysilicon backside remains challenging in the case of texturing[24,28]. Therefore, it is of value to assess how ultra-thin nanophotonic metagrating designs can tackle this challenge and selectively enhance the optical absorption while keeping the Si cell planar.

In order to maximize light trapping in the Si bottom cell, we optimize the periodicity $p$ of the hexagonal silver array acting as the back-reflector, along with height $h$ and radius $r$ of the nano-disk constituting its unit cell. The Ag disks are embedded in PMMA and form an integrated part of the Ag back contact of the solar cell. The figure-of-merit minimized is the sum of parasitic absorption occurring in the metal and back-reflection to the $0^{th}$ diffraction order at a single interaction, $\text{FOM} = \text{Abs} + R_{0^{th}}$ averaged in the bandwidth 1000-1200 nm. This, in turn, is equal to maximizing the fraction of light steered away at an angle, with the practical advantage of not having to consider a changing number of diffraction channels for every simulation. Figure 1a schematically shows the design parameters that are optimized. Employing a particle swarm optimization algorithm[29], the final set of retrieved optimal parameters is $p$=534 nm, $h$=240 nm, and $r$=171 nm (see Methods section). The corresponding spectra for parasitic absorption and reflection back to the $0^{th}$ diffraction order are shown in Fig. 1b. It is important to remark that reflection to the $0^{th}$-order is almost completely canceled while plasmonic absorption resonances are shifted outside the entire bandwidth of interest. The distribution of reflected power to the different diffraction channels (averaged over the simulation wavelength range and over both polarizations) is depicted in Fig. 1c. Most of the light reaching the back-reflector at the bottom of the Si cell is reflected equally to 6 diffraction channels beyond the critical angle for the Si/air interface with a ≈15.4% efficiency for each channel. This, in turn, implies that ≈92% of the incoming light is reflected at an angle within the TIR range. It is interesting to point out that, while the FOM is polarization independent, the coupling efficiencies to the channels at an angle are not (see Supplementary Information).

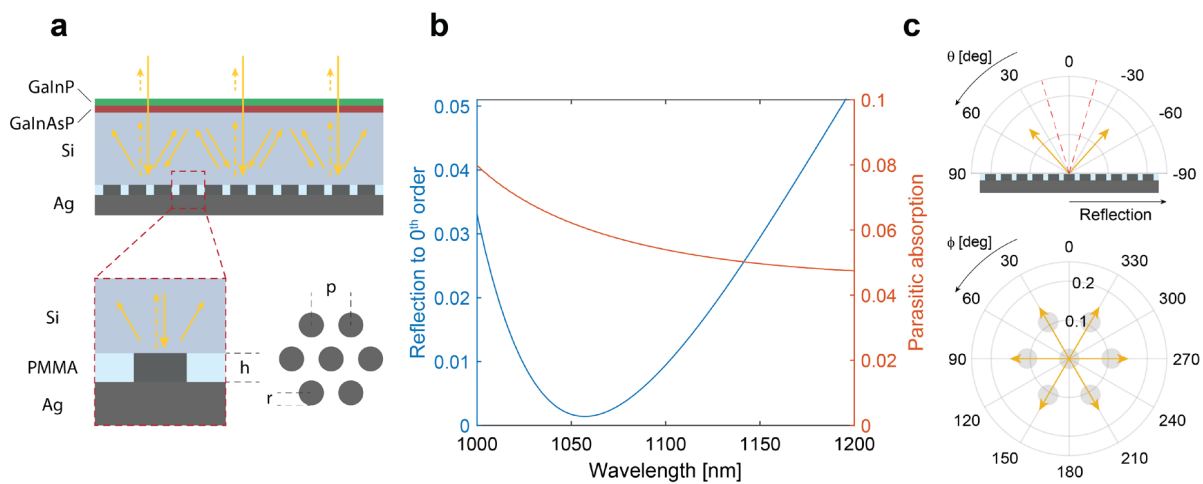

Fig. 1 | **Optimized back-reflector. a** Schematics of the GaInP/GaInAsP//Si multi-junction solar cell featuring a nanostructured Ag back-reflector at the Si sub-cell bottom. Back-reflector unit cell (side view and top view) highlighting the relevant design parameters optimized. **b** Reflection to the $0^{th}$ diffraction order (solid blue line) and parasitic absorption in the metal (solid orange line) for the optimized metagrating. **c** Polar plots showing the fraction of incident power reflected (radial coordinate) to each diffraction channel (angular coordinates $\theta$ and $\phi$) for unpolarized light averaged in the range 1000-1200 nm. The critical angle is indicated by the dashed red lines while the schematics indicates the grating position relative to the angular coordinates $\theta$ and $\phi$.

To understand why it is possible to achieve such a low FOM and to unravel the mechanism behind the back-reflector operation, it is useful to analyze the parameter space beyond the optimum values. Figure 2a-b shows the reflection to the $0^{th}$ diffraction order (a) and parasitic absorption (b), averaged in the range 1000-1200 nm, as a function of nano-disk height and periodicity if the lattice fill factor is kept constant ($p/r = 3$) and close to that of the optimal grating. As expected, if no diffraction channels are available the only way to have low reflection is by having a high parasitic absorption.

Indeed, for periodicities smaller than $p_{\min} = \lambda_{\max}/n_{\text{Si}} \approx 336$ nm the minima in Fig. 2a correspond to maxima in Fig. 2b. However, when additional diffraction channels are opened in the superstrate it becomes possible to find regions in the parameter space where both reflection and parasitic absorption are low.

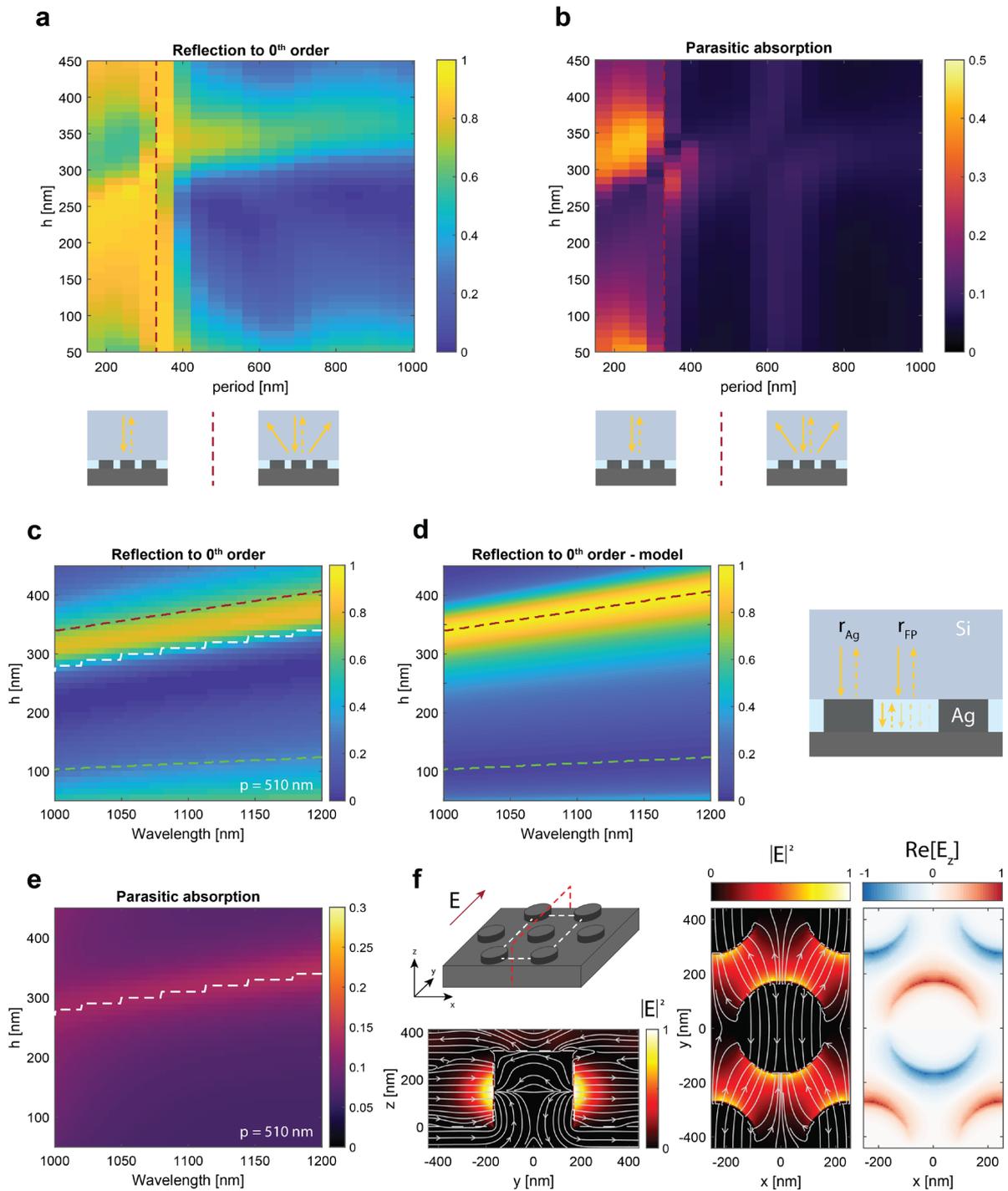

Fig. 2 | **Back-reflector design. a-b** Simulated reflection to the 0$^{\text{th}}$ diffraction order (a) and parasitic absorption in Ag (b), averaged in the range 1000-1200 nm, as a function of nano-disk height and periodicity if the lattice fill factor is kept constant ($p/r = 3$). The vertical red dashed lines indicate the onset of diffraction into Si, as schematically depicted in the bottom schematics. **c** FDTD simulated and (**d**) analytically modeled reflection to the 0$^{\text{th}}$ diffraction order as a function of

wavelength and nano-disk height for a fixed periodicity $p$ = 510 nm. The red and green dashed lines indicate respectively the conditions of maximum and the minimum reflection in (d). The dashed white line corresponds to the absorption maxima in (e). **e** Simulated parasitic absorption in the silver nano-disks as a function of wavelength and nano-disk height for a fixed periodicity $p$ = 510 nm and light at normal incidence. **f** Normalized electric field intensity and electric field $z$ component profiles on the cross-cut planes highlighted in the schematics. The height of the nano-disk is $h$ = 320 nm and the illumination wavelength is $\lambda$ = 1140 nm. The $x$-$y$ crosscut plane is taken at height $h$ = 160 nm corresponding to the middle of the nano-disk. This crosscut has a slight offset compared to the plane ($h \approx$ 150 nm) where the electric field $z$ component is zero for a dipole oscillating in-plane.

Next, taking crosscuts of the data in Fig. 2a-b at different periodicities it is possible to analyze the reflection to $0^{th}$-order and the parasitic absorption as a function of wavelength. Figure 2c-e shows such a crosscut for $p$ = 510 nm. Interestingly, the reflection to $0^{th}$-order (panel c) can be approximated with a simple model assuming that light that is reflected off the top of the nano-disks interferes with light reflected off the PMMA and Ag interfaces. While the first contribution can be captured by the Fresnel coefficient for the single Si/Ag interface, the second should take into account multiple reflection at the PMMA/Si and PMMA/Ag interfaces and thus be modeled as a Fabry-Pérot interference, as schematically shown in Fig. 2d (inset). Under these simple but insightful assumptions, the resulting Fresnel reflection coefficient for the entire back-reflector is

$$r_{\text{tot}} = F\, r_{\text{Ag}} + (1-F) r_{\text{FP}} \qquad (1)$$

where $F = \frac{2\pi r^2}{\sqrt{3} p^2}$ is the array fill factor, $r_{\text{Ag}}$ is the Fresnel reflection coefficient for the Si/Ag interface and $r_{\text{FP}}$ that of the PMMA etalon of height $h$ sandwiched between Si and Ag. Figure 2d shows the modeled reflection $|r_{\text{tot}}|^2$ as a function of wavelength and height. The trends in Fig. 2c are well reproduced in Fig. 2d but it is easy to notice that the conditions for minima and maxima (red and green dashed lines) are not matching exactly. Indeed, the simple model does not consider the plasmonic resonances in the Ag particle array. The presence of the latter is signaled by the resonant peak in parasitic absorption in Fig. 2e that redshifts for larger cylinder height. To further investigate its nature, we inspect the near field profiles at the resonant wavelength. Figure 2f shows the electric field profiles corresponding to a nano-disks array of height $h$ = 320 nm with an illumination wavelength $\lambda$ = 1140 nm. Looking at the electric field intensity distribution at different planes (see inset) as well as its component along the $z$ axis it can be seen that each nano-disk acts as a plasmonic dipole antenna that interacts with its neighbors via near-field coupling. This additional scattering pathway is interfering with non-resonantly reflected light, as described earlier and modeled with the coefficient $r_{\text{tot}}$. The interference between these *resonant* and *direct* pathways gives rise to a Fano lineshape in the reflection spectrum[23,30–32] where the resonance wavelength (white dashed line in Fig. 2c,e) is located between a maximum and a minimum in reflection, as shown in Fig. 2c.

It is worth pointing out that, different from other light redirection strategies employing gratings[33,34], in our case light is not steered at an angle by engineering the scattering of the single scattering unit via resonances. On the contrary, sharp resonant modes should be avoided as they would result in severe parasitic loss. Our design redirects light at an angle by suppressing one of the available channels ($0^{th}$ order) by destructive interference. Therefore, parasitic absorption in the metal can still be mitigated.

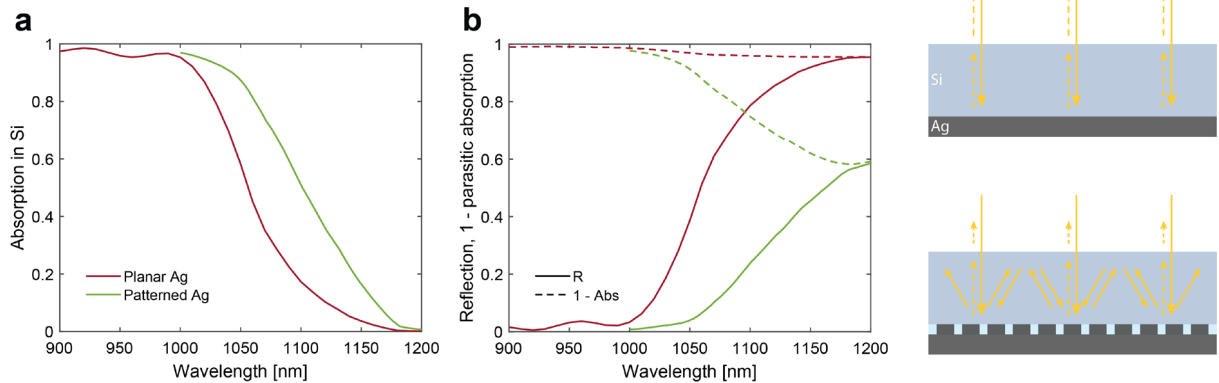

Fig. 3 | **Calculated absorption in the Si bottom cell: OPTOS-calculations. a** absorption spectrum near the Si band gap for the optimized design and for a cell with a planar Ag back-reflector. **b** reflection (solid lines) and (1- parasitic absorption) (dashed lines) spectra for planar (red curves) and patterned (green curves) cells. The calculations assume a double-layer anti-reflection coating and a Si cell thickness of 280 μm.

The analysis described so far concerns the light scattering from the Ag metagrating only. Next, we assess the benefits of this light trapping scheme on the entire solar cell. Although in principle possible, simulating the entire solar cell via FDTD is unfeasible from a computational point of view. Hence, the metagrating simulation/optimization and the calculation of the induced absorption enhancement within the thick Si bottom cell must be decoupled. One strategy to calculate the optical properties of optically thick sheets with arbitrarily textured front and rear surface is the OPTOS (Optical Properties of Textured Optical Sheets) formalism[35–37]. In this framework, the effect of the textured surface, in the present case the optimized metagrating, is captured by a redistribution scattering matrix where each element describes the fraction of light that is reflected from one diffraction channel to the other. In the scattering matrix formalism, the redistribution matrix is composed of the squared amplitudes of the s-parameters that constitute the complex-valued metagrating scattering matrix. The redistribution matrix is calculated independently via FDTD by simulating the grating response when light is incident from each diffraction channel, for each wavelength in the bandwidth 1000-1200 nm and both polarizations. The latter evaluation is rather time-consuming but has to be performed only once. Next, light propagation within the Si absorber is modeled with a propagation matrix that considers absorption according to Lambert-Beer's law. Thus, the optical properties of the entire stack (thick Si absorber and Ag grating) can be calculated by a series of matrix multiplications. The outcome of such calculation comparing the optimized design described above to a planar Ag back-reflector is shown in Fig. 3. The fraction of power absorbed in Si shows a clear enhancement due to the light trapping effect induced by the metagrating (Fig. 3a). This, in turn, results in a lower overall reflection compared to its planar counterpart (Fig. 3b). Even though parasitic absorption in the patterned Ag is higher, the increase in Si absorption still favors the metagrating design over the planar reference. Such analysis demonstrates that designing the metagrating at the nanoscale to suppress $0^{th}$-order reflection and parasitic absorption has a positive impact on the Si absorption on a hundreds-of-microns-scale.

**Experiment**

In order to experimentally demonstrate the proposed nanophotonic light trapping structure, large-scale nanopatterns are fabricated on solar cells by Substrate Conformal Imprint Lithography (SCIL)[38,39].

Full 4'' wafers with several silicon bottom cells or full GaInP/GaInAsP//Si triple-junction cells are fabricated with a planar Ag back-reflector and then characterized. Next, the planar Ag layer is peeled off mechanically, the front side is protected with resist and the nanopatterned back-reflector is fabricated on the clean Si back interface. The detailed fabrication procedure is explained step by step in the Methods section.

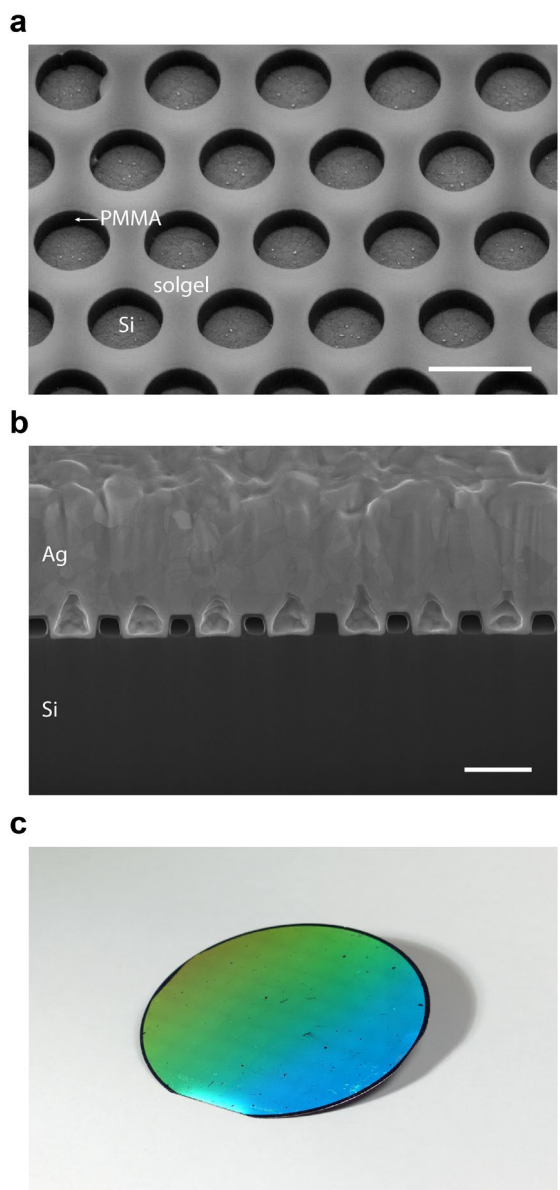

Fig. 4 | **Cell fabrication. a** Tilted SEM image of the patterned backside of the triple junction cell after the reactive ion etching steps (cell C1). **b** SEM image of a FIB cross-section of the same cell after metal deposition. The scale bar is 500 nm for panels a-b. **c** Photograph of the fully patterned backside of the wafer containing the triple-junction GaInP/GaInAsP//Si cells showing diffractive colors due to the fabricated metagrating.

Figure 4 shows the result of the fabrication procedure at two different stages. The SEM image (Fig. 4a) of one of the triple-junction cells (C1) after the reactive ion etching (RIE) steps indicates that the patterned solgel mask is smooth and uniform and the PMMA in the holes has low sidewall roughness. Also, no residual PMMA is left at the bottom of the holes after RIE. Figure 4b shows a cross section of the structure after Ag sputter deposition. As can be seen, conically shaped inclusions have formed in the metal film, which we attribute to non-perfectly straight metal deposition during the sputtering process. Nonetheless, such air inclusions do not deteriorate the optical performance of the back-reflector (see Supplementary Information). The samples are uniform over the entire patterned area with very few defects mainly due to dust particles (Fig. 4c).

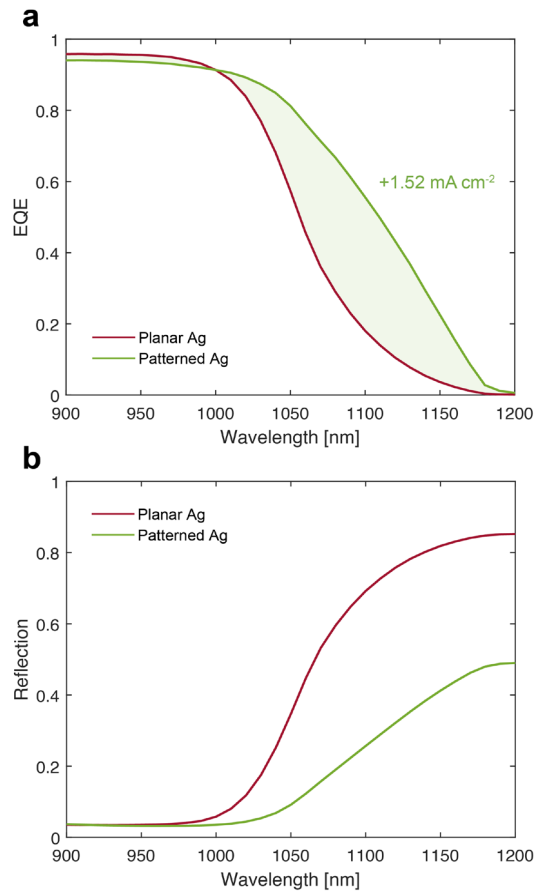

Fig. 5 | **Si bottom cell performance.** Experimental **a** EQE and **b** reflection spectra for a Si solar cell with the optimized design described above and for a cell with a planar Ag back-reflector. The Si cell had a thickness of 300 μm and a double anti-reflection coating was applied.

To assess the impact of patterning on a device level, silicon bottom cells, and full two-terminal triple-junction cells are optically and electronically characterized, first with a planar Ag back-reflector and finally with the fabricated optimized grating. Starting with Si bottom cells, Fig. 5a highlights the enhancement in the External Quantum Efficiency (EQE) spectra due to the back-reflector nano-structuring. As anticipated by simulation, the experimental spectra demonstrate a distinct advantage in using the described optimized design. The lower reflection achieved by the hexagonal grating (see Fig. 5b) can be ascribed to enhanced light trapping capabilities and parasitic absorptance. The boost in EQE results in a +1.52 mA/cm$^2$ current density gain. To the best of our knowledge this is the highest gain measured in this type of cells due to back-reflector nanostructuring. These experimental results fully corroborate the light trapping metagrating design and modeling approach described above.

Given the performance improvement validated on Si bottom cells, full two-terminal triple-junction GaInP/GaInAsP//Si cells were patterned and characterized. Figure 6a depicts the front side of the wafer containing cell C1, showing the cell configuration layout, and the patterned backside of another sample.

The EQE spectra of cell C1 are shown in Fig. 6b for a planar back reflector. As calculated from these spectra, the $J_{SC}$ of each sub-sell indicates that the Si cell is limiting the current flow of the entire stack. Hence, any current gain in the Si sub-cell directly improves the $J_{SC}$ of the entire device.

Similar to the observation in Fig. 5, also in this case, there is a clear increase in EQE when the planar Ag reflector is replaced with the optimized metagrating (see Fig. 6c). The current density gain calculated from the EQE (data in Fig. 6c) amounts to 1.22 mA/cm$^2$. It is important to stress that the very same cell is measured first with a planar Ag and, once the back-reflector is peeled off and a new patterned one is fabricated, with the optimized Ag grating design.

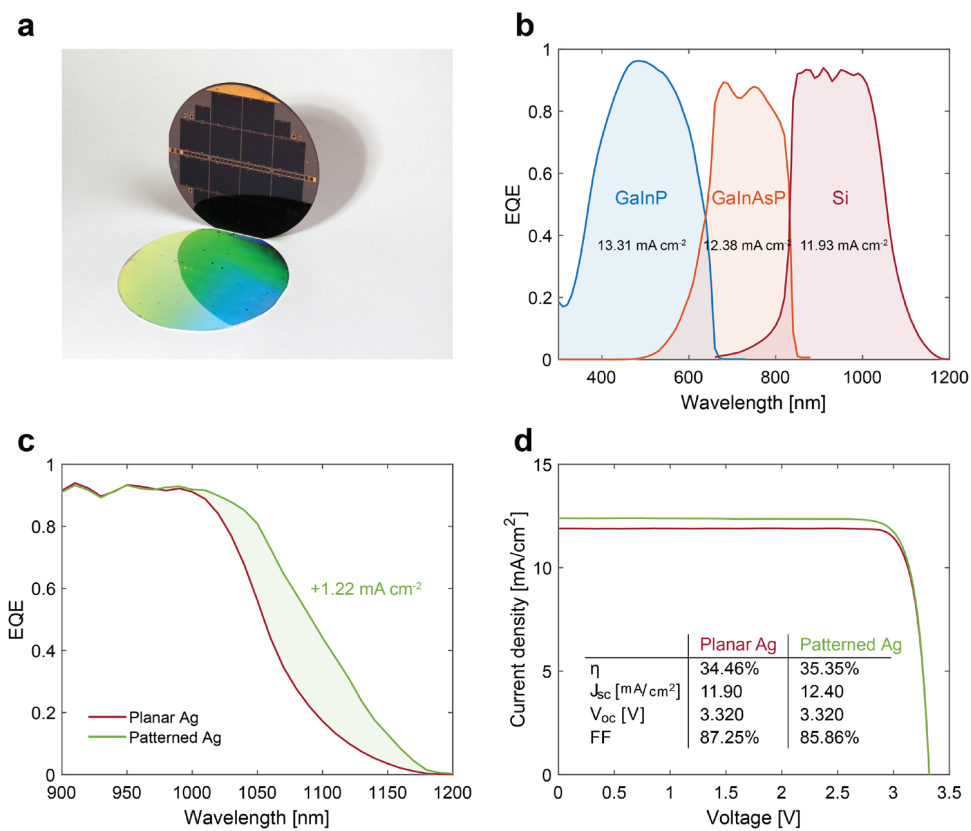

Fig. 6| **Full triple-junction cell performance. a** Photograph of the fully patterned backside of the 4'' wafer containing all triple-junction cells. **b** Experimental EQE spectra of each sub-cell with a planar electrode for cell C1. **c** Experimental EQE spectra of the Si sub-cell for a patterned and planar back-reflector. **d** Measured one-sun I-V characteristics comparing the cell with the optimized hexagonal grating and its planar reference. All panels refer to the same triple-junction cell C1 with either a planar or a patterned back-contact.

Calibrated one-sun current-voltage measurements were performed on cell C1 at ISE CalLab and the power conversion efficiency $\eta$, short circuit current density $J_{OC}$, open circuit voltage $V_{OC}$, and fill factor *FF* were determined; the data are summarized in the inset of Fig. 6d. Corroborating the EQE enhancement, the I-V curves prove a clear increase in the $J_{SC}$ due to optimized light trapping and consequently higher absorption in Si. Importantly, the short-circuit current gain for the entire stack amounts to 0.5 mA/cm$^2$. This value is lower compared to the estimated gain in the Si sub-cell (1.22 mA/cm$^2$, see above) as the GaInAsP middle cell is now limiting the overall current flow in the series-connected device. We note that the $V_{OC}$ of the patterned cell remains essentially unaltered compared to the planar reference. This implies that the cells are not damaged or electronically degraded by the back-reflector processing, further proving the applicability of SCIL to the fabrication of high-efficiency solar cells. The light trapping effect induced by the implemented metagrating design results in an efficiency increase of +0.9 % (absolute) for cell C1, reaching an overall efficiency of 35.35 %.

Additional measurements of cells on the same wafer and of other wafers, all patterned using the same SCIL stamp, show small variations in efficiency, with a highest efficiency observed of 35.6% and a highest Si sub-cell current gain of 1.52 mA/cm$^2$ (data shown in Supplementary Information).

**Conclusion**

In conclusion, the results presented in this work demonstrate a nanophotonic light trapping scheme for Si-based multi-junction solar cells that integrates and optimizes near-field nanophotonic light scattering from a metallo-dielectric metagrating with a far-field multiple scattering matrix formalism. The scattering matrix is composed of elements derived from near-field simulations and a far-field scattering matrix multiplication analysis leads to the evaluation of light trapping in the silicon bottom cell.

We introduce the concept of a diffractive metagrating back-reflector composed of a hexagonal array of silver nano-disks embedded in PMMA and integrated within the silver back contact. The nano-disk radius and height as well as the lattice periodicity are optimized by employing a particle swarm optimization algorithm. The optimal design shows minimal back-reflection to the 0$^{th}$ diffraction order while keeping parasitic plasmonic absorption low. The working principle of the design can be captured by a simple interference model. We then propagate the optimized nanoscale design to the macroscale via the OPTOS multiple-scattering formalism, to evaluate the absorption in a thick Si slab, highlighting a clear advantage of the metagrating over a planar back-reflector.

The resulting patterned Si bottom cells and full two-terminal GaInP/GaInAsP//Si triple-junction cells are characterized optically and electronically. We show experimentally a +0.9 % (absolute) efficiency gain on full triple-junction cells and superior EQE on Si bottom cells (+1.52 mA/cm$^2$). We note that the integration of a nanopatterned surface in the process flow of a high-efficiency cell proved to be possible via SCIL with no signs of degradation of the semiconductor quality.

Altogether, these results show a nanophotonic light trapping scheme based on metagratings that can enhance the performance of cells where more traditional random texture strategies are not possible, such as the wafer-bonded TOPCon-based cells in this work. The results may also be relevant for the upcoming research field of perovskite/Si tandem solar cells, where a planar Si surface is often beneficial for the fabrication of a high-quality perovskite top cell.

**Methods**

**Numerical simulations**

All optical simulations were performed using the software FDTD: 3D Electromagnetic Simulator by Lumerical Inc.

To optimize the back reflector geometry (data in Fig.1), two perfectly matched layers (PMLs) have been used for the top and the bottom of the simulated region while periodic boundary conditions were imposed in plane. The array of Ag cylinders is placed on a semi-infinite Ag substrate while the Si semi-infinite superstrate is modeled as a lossless material with a refractive index *n* = 3.57. A plane wave source is used to inject light from the Si side at normal incidence. The reflection to the 0$^{th}$ diffraction order is extracted via the "grating order transmission" analysis group while the parasitic absorption is calculated employing a "power transmission box" analysis group surrounding the pattern. The sum of these two quantities is used as a figure of merit (FOM) to be minimized within the software's Particle Swarm Optimization (PSO) utility. The parameters optimized are the periodicity *p*, height *h* and radius *r* of the nanocylinders. To reduce the parameter space explored by the optimization, the nano-disk radius and height are constrained to practical values (200 nm < *h* < 400 nm, 50 nm < *r* < 450 nm) taking into account boundary conditions in fabrication. The range of periodicities explored is 300 nm < *p* < 1200 nm, slightly larger than $p_{min} < p < p_{max}$, where $p_{min} = \lambda_{max}/n_{Si} \approx 336$ nm is the minimum periodicity required to have diffraction in the entire bandwidth and $p_{max} = (2/\sqrt{3})\lambda_{min} \approx 1154$ nm is the maximum periodicity for which the first diffraction orders are outside the escape cone (≈16.2° for the Si/air interface). Different local minima of the FOM can be found with such constraints, the range 300 nm < *p* < 600 nm was used to obtain the optimum parameters used in this work. A similar simulation configuration is used to explore the parameter space, as shown in Fig. 2.

To calculate the redistribution matrices needed to perform OPTOS calculation, the simulation featuring the optimum parameters illuminated from normal incidence is used as a seed to set up all the other simulations characterizing the grating response when light is incident from each diffraction channel, for each wavelength, and for both polarizations. For these simulations, Bloch boundary conditions are employed.

**Cell Fabrication**

The III-V top cells were grown in a commercial AIXTRON AIX2800G4-TM reactor using metalorganic vapour-phase epitaxy (MOVPE). Epitaxial growth was performed on 4" GaAs wafers with an offcut of the (001) surface towards [111]B at temperatures ranging from 550 °C to 640 °C. More details on the epitaxial growth are reported elsewhere[24,27].

Silicon bottom cells were fabricated on polished p-type float-zone silicon wafers with bulk resistivity of 1 Ωcm and thickness of 300 µm. Tunnel-oxide passivating contacts (TOPCon)[40] were formed on both sides with n-type doping on the front and p-type doping on the back side. The thin tunnel oxide was grown thermally in a tube furnace at 600 °C. A 100 nm amorphous silicon layer was deposited using LPCVD, doped by ion implantation, and crystalized at 900 °C. The passivation was improved by a remote hydrogen plasma at 425 °C. More details of the silicon bottom cells can be found elsewhere[24].

The upright-grown III-V top cells were temporarily bonded to a sapphire carrier wafer and the GaAs substrate was removed to bare the bond layer. The III-V bond layer was cleaned and thinned by chemical-mechanical polishing (CMP), but the silicon bottom cells were only cleaned without abrasion.

The top and bottom cells were joined by direct wafer bonding at room temperature, with surface activation by an argon beam. After bonding, the sapphire carrier wafer was removed by thermal slide at 190 °C. Solar cells with a size of 2 cm x 2 cm were fabricated including front contacts, anti-reflection coating (ARC), and a 1 µm thick Ag layer on the back side acting as electrical contact and planar mirror. More details on the cell processing are published elsewhere[24,27].

**Back-reflector nanopatterning**

The silicon master used to mold the double-layer PDMS stamp was acquired from NIL Technology. It consists of a 10 cm$^2$ array of holes with the optimized periodicity and radius described above and a depth of about 122 nm which is optimal for solgel imprinting and PDMS demolding. Silicon bottom cells and full triple-junction cells were fabricated at Fraunhofer ISE with a planar Ag back-reflector and then characterized. Next, the planar Ag was peeled off mechanically, the front side was protected with AZ520D resist and the cells were shipped to AMOLF for patterning. After nanoimprint and etching the cells were shipped again to Fraunhofer ISE for a final HF dip followed by metallization. In the following, the detailed fabrication procedure is explained step by step:

(1) A layer of PMMA (PMMA 950k A8, 240 nm) is spin-coated and baked at 150˚C for 2 minutes. The thickness of the layer is equal to the final Ag nano-disk height.

(2) A layer of solgel (SCIL Nanoimprint Solutions NanoGlass T-1100, 75 nm) is spin-coated and, before solidification, the PDMS stamp is slowly brought in contact to mold the layer. After 6 minutes of curing time, the stamp is detached carefully.

(3) The solgel residual layer at the bottom of the imprinted holes is cleared via RIE using Oxford Instrument's PlasmaPro 80 and a process employing $CHF_3$ and Ar (etch rate 0.28 nm/s).

(4) The PMMA exposed in the solgel holes is etched down completely via Reactive Ion Etching (RIE) using an $O_2$ plasma (etch rate 7.94 nm/s).

(5) The sample is shipped back to Fraunhofer ISE for further processing: a wet HF etch (5 min) removes the patterned solgel mask and the native oxide on the exposed Si surface inside the PMMA holes. This ensures proper electrical contact. Finally, 1 µm Ag is sputtered and the front resist protection is removed.

**Device characterization**

External Quantum Efficiencies (EQEs) were measured using a grating monochromator set-up with adjustable bias voltage and bias spectrum[41] while reflection measurements were performed on an integrated LOANA measurement device. One-sun I–V characteristics were measured under a spectrally adjustable solar simulator[42] with one xenon lamp and two halogen lamp fields that were adjusted in intensity independently of each other to generate the same current densities in each sub-cell as under illumination with the AM1.5G spectrum (IEC 90604-3, ed. 2 with 1000 W/m$^2$)[42]. The cell temperature was held at 25 ˚C during the measurement. An aperture mask with an area of 3.984 cm$^2$ was placed on top of the III–V/Si solar cell to avoid any contribution of photo-generated carriers from outside the defined cell area.

**Acknowledgments**

This work is part of the research program of the Dutch Research Council (NWO). It was partly funded by the Fraunhofer Gesellschaft (ICON project MEEt).

The authors thank the Fraunhofer ISE employees R. Koch, R. Freitas, P. Simon, A. Lösel, A. Leimenstoll, F. Schätzle, F. Martin, M. Schachtner, E. Fehrenbacher, D. Chojniak, F. Sahajad, S. Stättner, D. Lackner, G. Siefer, and F. Dimroth for their support in device processing, characterization, and helpful discussions.

**Author contributions**

AC conceived the back reflector design by performing numerical simulations and theoretical analyses which were supported by NT and BB. RM, PS and OH fabricated and characterized the solar cells (before and after patterning). AC and SWT performed nanofabrication and structural characterization. BB, OH and AP supervised the project.

All authors contributed to the analysis and writing of the paper.

**Competing interests**

The authors declare no competing interests.

**Data availability**

The data that support the findings of this study are available from the corresponding author upon reasonable request.

**Code availability**

All codes produced during this research are available from the corresponding author upon reasonable request.

**ORCID**

Albert Polman 0000-0002-0685-3886

Andrea Cordaro 0000-0003-3000-7943

Ralph Müller 0000-0001-6248-3659

Oliver Höhn 0000-0002-5991-2878

Patrick Schygulla 0000-0001-9103-1045

Benedikt Bläsi 0000-0003-1624-1530

**References**

1. IEA. International Energy Agency (IEA) World Energy Outlook 2022. *World Energy Outlook 2022* (2022). doi:https://www.iea.org/reports/world-energy-outlook-2022

2. Polman, A., Knight, M., Garnett, E. C., Ehrler, B. & Sinke, W. C. Photovoltaic materials: Present efficiencies and future challenges. *Science (80-. ).* **352**, 307 (2016).

3. Ehrler, B. *et al.* Photovoltaics Reaching for the Shockley–Queisser Limit. *ACS Energy Lett.* **5**, 3029–3033 (2020).

4. NREL. Best Research-Cell Efficiency Chart. (2022). Available at: https://www.nrel.gov/pv/cell-efficiency.html.

5. Ho-Baillie, A. W. Y. *et al.* Recent progress and future prospects of perovskite tandem solar cells. *Appl. Phys. Rev.* **8**, (2021).

6. Yamaguchi, M., Dimroth, F., Geisz, J. F. & Ekins-Daukes, N. J. Multi-junction solar cells paving the way for super high-efficiency. *J. Appl. Phys.* **129**, 240901 (2021).

7. Yoshikawa, K. *et al.* Silicon heterojunction solar cell with interdigitated back contacts for a photoconversion efficiency over 26%. *Nat. Energy* **2**, 17032 (2017).

8. Smith, D. D. *et al.* Toward the Practical Limits of Silicon Solar Cells. *IEEE J. Photovoltaics* **4**, 1465–1469 (2014).

9. Richter, A., Hermle, M. & Glunz, S. W. Reassessment of the Limiting Efficiency for Crystalline Silicon Solar Cells. *IEEE J. Photovoltaics* **3**, 1184–1191 (2013).

10. Shockley, W. & Queisser, H. J. Detailed Balance Limit of Efficiency of p-n Junction Solar Cells. *J. Appl. Phys.* **32**, 510–519 (1961).

11. Green, M. A. & Bremner, S. P. Energy conversion approaches and materials for high-efficiency photovoltaics. *Nat. Mater.* **16**, 23–34 (2017).

12. Zhu, J. *et al.* Optical Absorption Enhancement in Amorphous Silicon Nanowire and Nanocone Arrays. *Nano Lett.* **9**, 279–282 (2009).

13. Atwater, H. A. & Polman, A. Plasmonics for improved photovoltaic devices. *Nat. Mater.* **9**, 205–213 (2010).

14. Cordaro, A. *et al.* Nanopatterned SiNx Broadband Antireflection Coating for Planar Silicon Solar Cells. *Phys. status solidi* **2200827**, 2200827 (2023).

15. Garnett, E. C., Ehrler, B., Polman, A. & Alarcon-Llado, E. Photonics for Photovoltaics: Advances and Opportunities. *ACS Photonics* **8**, 61–70 (2021).

16. Yu, Z., Raman, A. & Fan, S. Fundamental limit of nanophotonic light trapping in solar cells. *Proc. Natl. Acad. Sci.* **107**, 17491–17496 (2010).

17. Zhu, J., Hsu, C. M., Yu, Z., Fan, S. & Cui, Y. Nanodome solar cells with efficient light management and self-cleaning. *Nano Lett.* **10**, 1979–1984 (2010).

18. Spinelli, P., Verschuuren, M. & Polman, A. Broadband omnidirectional antireflection coating based on subwavelength surface Mie resonators. *Nat. Commun.* **3**, 692 (2012).

19. Brongersma, M. L., Cui, Y. & Fan, S. Light management for photovoltaics using high-index nanostructures. *Nat. Mater.* **13**, 451–460 (2014).

20. van de Groep, J., Spinelli, P. & Polman, A. Transparent Conducting Silver Nanowire Networks. *Nano Lett.* **12**, 3138–3144 (2012).

21. Van De Groep, J., Spinelli, P. & Polman, A. Single-Step Soft-Imprinted Large-Area

Nanopatterned Antireflection Coating. *Nano Lett.* **15**, 4223–4228 (2015).

22. Pecora, E. F., Cordaro, A., Kik, P. G. & Brongersma, M. L. Broadband Antireflection Coatings Employing Multiresonant Dielectric Metasurfaces. *ACS Photonics* **5**, 4456–4462 (2018).

23. Cordaro, A. *et al.* Antireflection High-Index Metasurfaces Combining Mie and Fabry-Pérot Resonances. *ACS Photonics* **6**, 453–459 (2019).

24. Cariou, R. *et al.* III–V-on-silicon solar cells reaching 33% photoconversion efficiency in two-terminal configuration. *Nat. Energy* **3**, 326–333 (2018).

25. Tillmann, P. *et al.* Optimizing metal grating back reflectors for III-V-on-silicon multijunction solar cells. *Opt. Express* **29**, 22517 (2021).

26. Bläsi, B., Hanser, M., Jäger, K. & Höhn, O. Light trapping gratings for solar cells: an analytical period optimization approach. *Opt. Express* **30**, 24762 (2022).

27. Schygulla, P. *et al.* Two-terminal III–V//Si triple-junction solar cell with power conversion efficiency of 35.9 % at AM1.5g. *Prog. Photovoltaics Res. Appl.* 1–11 (2021). doi:10.1002/pip.3503

28. Larionova, Y. *et al.* On the recombination behavior of p + -type polysilicon on oxide junctions deposited by different methods on textured and planar surfaces. *Phys. status solidi* **214**, 1700058 (2017).

29. Lumerical Inc. FDTD: 3D Electromagnetic Simulator.

30. Fano, U. Sullo spettro di assorbimento dei gas nobili presso il limite dello spettro d'arco. *Nuovo Cim.* **12**, 154–161 (1935).

31. Ruan, Z. & Fan, S. Temporal Coupled-Mode Theory for Fano Resonance in Light Scattering by a Single. *J. Phys. Chem. C* 7324–7329 (2010). doi:10.1021/jp9089722

32. Haus, H. A. & Huang, W. Coupled-Mode Theory. *Proc. IEEE* **79**, 1505–1518 (1991).

33. Khaidarov, E. *et al.* Asymmetric Nanoantennas for Ultrahigh Angle Broadband Visible Light Bending. *Nano Lett.* **17**, 6267–6272 (2017).

34. Ra'di, Y., Sounas, D. L. & Alù, A. Metagratings: Beyond the Limits of Graded Metasurfaces for Wave Front Control. *Phys. Rev. Lett.* **119**, 067404 (2017).

35. Eisenlohr, J. *et al.* Matrix formalism for light propagation and absorption in thick textured optical sheets. *Opt. Express* **23**, A502 (2015).

36. Tucher, N. *et al.* 3D optical simulation formalism OPTOS for textured silicon solar cells. *Opt. Express* **23**, A1720 (2015).

37. Tucher, N. *et al.* Optical simulation of photovoltaic modules with multiple textured interfaces using the matrix-based formalism OPTOS. *Opt. Express* **24**, A1083 (2016).

38. Verschuuren, M. A., Megens, M., Ni, Y., van Sprang, H. & Polman, A. Large area nanoimprint by substrate conformal imprint lithography (SCIL). *Adv. Opt. Technol.* **6**, 243–264 (2017).

39. Verschuuren, M. A., Knight, M. W., Megens, M. & Polman, A. Nanoscale spatial limitations of large-area substrate conformal imprint lithography. *Nanotechnology* **30**, 345301 (2019).

40. Feldmann, F., Bivour, M., Reichel, C., Hermle, M. & Glunz, S. W. Passivated rear contacts for high-efficiency n-type Si solar cells providing high interface passivation quality and excellent transport characteristics. *Sol. Energy Mater. Sol. Cells* **120**, 270–274 (2014).

41. Siefer, G., Gandy, T., Schachtner, M., Wekkeli, A. & Bett, A. W. Improved grating monochromator set-up for EQE measurements of multi-junction solar cells. *Conf. Rec. IEEE Photovolt. Spec. Conf.* 86–89 (2013). doi:10.1109/PVSC.2013.6744105

42. Meusel, M., Adelhelm, R., Dimroth, F., Bett, A. W. & Warta, W. Spectral mismatch correction and spectrometric characterization of monolithic III-V multi-junction solar cells. *Prog. Photovoltaics Res. Appl.* **10**, 243–255 (2002).